\documentclass[preprint,prb,showpacs]{revtex4}

\usepackage{graphicx}
\usepackage{dcolumn}
\usepackage{float}
\usepackage{amsmath}

\newcommand{\comment}[1]{}

\begin{document}

\title{\boldmath Evolution of electronic structure of Ru-doped single-crystal iridiates, Sr$_2$Ir$_{1-x}$Ru$_x$O$_4$ \unboldmath}

\author{Seokbae Lee$^{1}$}
\thanks{These authors contributed equally.}
\author{Yu-Seong Seo$^{1}$}
\thanks{These authors contributed equally.}
\author{Eilho Jung$^1$} \author{Seulki Roh$^1$} \author{Myounghoon Lee$^1$} \author{Hwan Young Choi$^2$} \author{Jong Hyuk Kim$^2$} \author{Nara Lee$^2$} \author{Young Jai Choi$^2$} \author{Jungseek Hwang$^{1}$}\email{jungseek@skku.edu}

\affiliation{$^{1}$Department of Physics, Sungkyunkwan University, Suwon, Gyeonggi-do 16419, Republic of Korea \\ $^{2}$Institute of Physics and Applied Physics, Yonsei University, Seoul 03722, Republic of Korea}

\date{\today}

%
%
\begin{abstract}

We investigated Ru-doped single-crystal 5$d$ iridiates, Sr$_2$Ir$_{1-x}$Ru$_x$O$_{4}$, at three different doping concentrations ($x =$ 0.01, 0.07 and 0.10) using optical spectroscopy. The undoped pristine compound (Sr$_2$IrO$_{4}$) is known as a novel $J_{eff}$ = 1/2 Mott insulator. Remarkably, the optical conductivity spectra of all three samples exhibited the insulating behavior, although we observed weak Drude components in the optical conductivity spectra down to the lowest temperature of 30 K. The charge-carrier densities of the Ru-doped iridiates estimated from the Drude components are significantly smaller than the expected values estimated from the nominal Ru-doping concentrations. Herein, we provide temperature- and doping-dependent electronic structure evolution of Ru-doped iridiates. We expect that our results will be useful for understanding the intriguing Ru-doping-dependent properties of 5$d$ iridiate Sr$_2$IrO$_{4}$.

\end{abstract}

\pacs{78.20.-e, 78.30.-j, 78.40.-q}

\maketitle


\section{Introduction}
A 5$d$ ternary compound, Sr$_2$IrO$_{4}$, was discovered many years ago\cite{randall:1957}. Since then, 5$d$ transition metal (iridium) oxides have been extensively investigated\cite{cava:1994,moon:2008,kim:2008,moon:2009,kim:2009a,lee:2012,ye:2013,sohn:2014,clancy:2014,cao:2016}. Because the 5$d$ transition metal oxides exhibit a significantly larger bandwidth than that of 3$d$ metal oxides, the Coulomb interaction is reduced and the spin-orbit coupling enhanced. These differences may spawn interesting new electronic ground states of 5$d$ transition metal oxides because of the three competing energy scales of bandwidth ($W$), Coulomb repulsion ($U$), and spin-orbit coupling ($E_{SO}$). Appropriate doping can be used to control these electronic tuning parameters\cite{lee:2012}. By replacing 5$d$ Ir atom with 4$d$ Ru, the band-filling (hole doping), $E_{SO}$, and the ratio of the Coulomb repulsion to the bandwidth ($U/W$) through structural distortion ($W$) and orbit extension difference ($U$) can be tuned\cite{subramanian:1994,lee:2012}. An optical spectroscopic study on Ru-doped Sr$_2$IrO$_4$ (Sr-214) thin films at room temperature was performed and it provided interesting doping-dependent electronic evolutions\cite{lee:2012}. In a recent optical study on La-doped Sr-214 ((Sr$_{1-x}$La$x$)$_2$IrO$_4$) single crystals, pesudogap effects were observed below the onset temperature of short-range antiferromagnetic order and the anomalous charge dynamics were attributed to the electronic correlations\cite{seo:2017}. In another optical study on La-doped Sr-214 single crystals, a soft collective mode and a Drude mode were observed simultaneously in over 10\% La-doped Sr-214 and part of its Fermi surface was ungapped at low temperatures\cite{wang:2018}. Other studies have been performed on Rh-doped Sr-214\cite{clancy:2014,cao:2016}. One may expect that these Rh-doped Sr-214 systems are isovalently doped. However, studies revealed that Rh-doped Sr-214 became hole-doped owing to the introduction of Rh$^{3+}$/Ir$^{5+}$. An intriguing question is whether doped Sr-214 can be used as a platform for superconductivity. A study on angle-resolved photoemission spectroscopy observed a gap with the $d$-wave symmetry in electron-doped Sr-214\cite{torre:2015,kim:2016}. However, superconductivity has not been clearly observed in doped Sr-214. A Raman study on Ru-doped Sr-214 systems revealed that a small percentage of Ru-doping had a strong effect on spin excitations and lattice dynamics\cite{glamazda:2014}. An optical spectroscopy study on Ru-doped Sr-214 single crystals may provide interesting new information, particularly on Ru-doping-dependent evolutions of charge-carrier density and electronic structure.

In this study, we investigated Ru-doped Sr-214 single crystals, Sr$_2$Ir$_{1-x}$Ru$_x$O$_{4}$ ($x =$ 0.01, 0.07, and 0.10), using optical spectroscopy. Optical spectroscopy is a suitable experimental technique for studying the charge-carrier dynamics, electronic ground state, and electronic structures of a particular material system. We expected that the hole-doping level of Sr-214 could be controlled by Ru-doping because the 4$d$ Ru atom has one less electron than the 5$d$ Ir atom. However, our results indicate that the doping effect is negligibly small in terms of charge-carrier density. The charge-carrier densities obtained from measured optical spectra were significantly smaller than those estimated using the energy dispersive spectrometry (EDS) technique. However, we obtained the Ru-doping-dependent electronic band structure evolution of Sr-214 from the results of Drude-Lorentz model analysis\cite{wooten,tanner:2019} and discussed about the electronic band structure evolution.

\section{Experiments}

\begin{figure}[!htbp]
  \vspace*{-0.3 cm}%
  \centerline{\includegraphics[width=4.5 in]{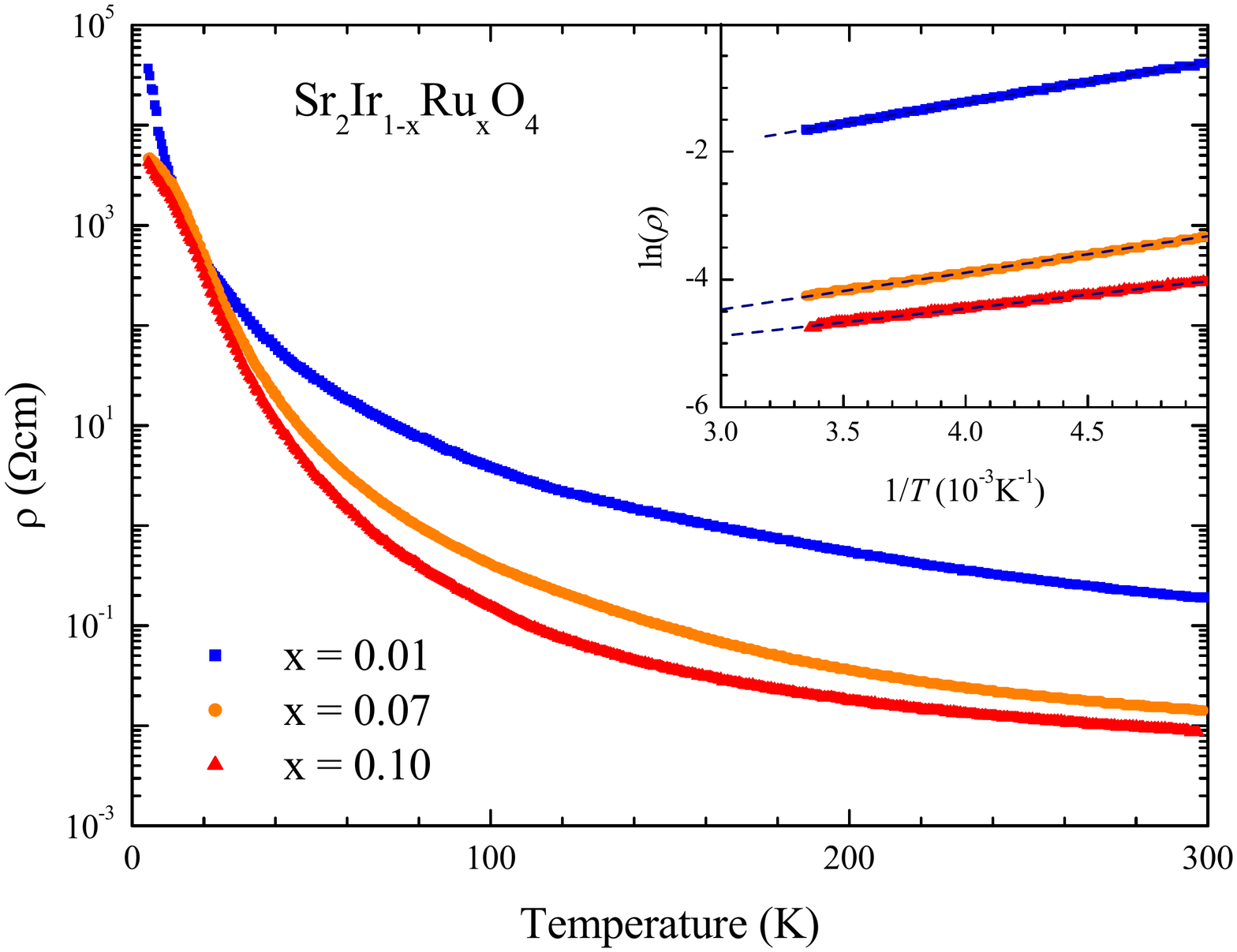}}%
  \vspace*{-0.7 cm}%
\caption{(Color online) Measured DC resistivity data of Sr$_2$Ir$_{1-x}$Ru$_x$O$_4$ in three different Ru-doping levels of $x$ = 0.01, 0.07, and 0.10. The Arrhenius plots are shown in the inset. The dashed lines are linear-fits. }
 \label{fig0}
\end{figure}

High-quality single crystals of Sr$_2$Ir$_{1-x}$Ru$_x$O$_4$ ($x$ = 0.01 , 0.07, and 0.10) were grown using a self-flux method\cite{kim:2009a}. To verify Ru concentrations of Sr$_2$Ir$_{1-x}$Ru$_x$O$_4$ crystals, the energy dispersive spectrometry (EDS) technique was employed using a JEOL-7800F (JEOL Ltd.) field emission scanning electron microscope equipped with an energy dispersive x-ray spectrometer working at 5.0 kV. The estimated Ru concentrations were 0.018$\pm$0.002, 0.070$\pm$0.005, and 0.104$\pm$0.011, respectively, for the samples with $x =$ 0.01, 0.07 and 0.10. The sample with $x =$ 0.01 shows a larger concentration than the nominal one. DC resistivity data were measured using a conventional 4-point probe technique and a physical property measurement system (PPMS, Quantum Design, Inc.). The measured DC resistivity data of the all three samples are shown in Fig. \ref{fig0}, which interestingly exhibit insulating temperature-dependent behavior. In the inset, we show the Arrhenius plots of the DC resistivity data of the three samples in a temperature range from 300 to 200 K. From the Arrhenius plot, we estimated the activation energy $E_a$. The estimated $E_a$ are 55, 50, and 37 meV for the samples with $x =$ 0.01, 0.07, and 0.10, respectively. The prepared single crystals were considerably small ($<$ 1 mm$^2$ area) and brittle. The size of the sample became smaller as the doping increased. However, because the surfaces of all samples were flat and bright, we could obtain reliable spectra. Because the size of Ru ion is similar to that of Ir ion, the unit cell volume with respect to Ru-doping remains within 0.5\% up to 10\% Ru-doping concentrations \cite{cava:1994}. Therefore, all three samples had the same crystal symmetry: a tetragonal symmetry (space group $I4_1$/a)\cite{torchinsky:2015}.

\begin{figure}[!htbp]
  \vspace*{-0.3 cm}%
  \centerline{\includegraphics[width=4.5 in]{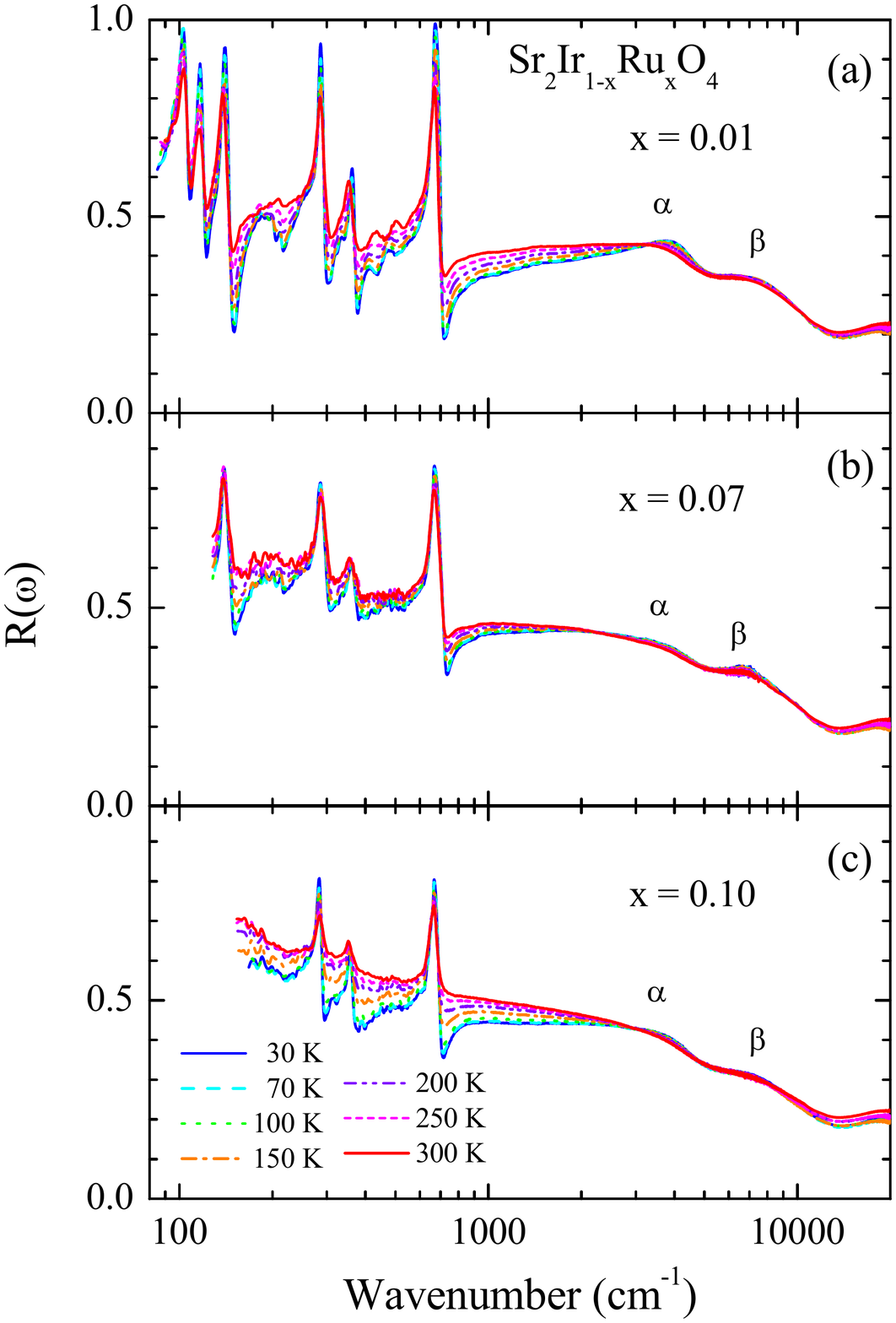}}%
  \vspace*{-0.7 cm}%
\caption{(Color online) Measured reflectance spectra of Sr$_2$Ir$_{1-x}$Ru$_x$O$_4$ in three different Ru-doping levels at various selected temperatures: (a) $x$ = 0.01, (b) $x$ = 0.07, and  (c) $x$ = 0.10.}
 \label{fig1}
\end{figure}

We measured near-normal reflectance spectra in the spectral range from 80 to 13,000 cm$^{-1}$ at several selected temperatures from 30 to 300 K using a commercial Fourier-transform infrared spectrometer (Bruker Vertex 80v) and a continuous liquid helium flow cryostat. To obtain accurate reflectance spectra in a wide spectral range up to 20,000 cm$^{-1}$, we employed an {\it in-situ} gold/aluminum evaporation technique\cite{homes:1993}. We obtained reflectance spectra of the sample with $x=$ 0.01, which was the largest sample, from 8000 to 20,000 cm$^{-1}$ at the selected temperatures, and used the measured reflectance spectra to extend the data up to 20,000 cm$^{-1}$ for all three samples. We note that, for samples with $x =$ 0.07 and 0.10, the extensions to 20,000 cm$^{-1}$ were approximates; we used them just for the Kramers-Kronig analysis. The measured reflectance spectra of the three samples at various selected temperatures are shown in Fig. \ref{fig1}. Because the size of the sample became smaller with increasing the doping the low-frequency cutoff of the measured spectra was $\sim$150 cm$^{-1}$ for the highest doped (10\%) sample due to the diffraction limit. As we expect from the DC resistivity data, all three samples exhibited insulating temperature-dependent behavior; the reflectance at the low-frequency region below 2000 cm$^{-1}$ was enhanced as the temperature increased. The overall reflectance was enhanced as the doping increased. However, the doping-dependent enhancement of reflectance at low-frequency region is not very dramatic, comparing to that of 3$d$ transition metal oxides such as cuprates\cite{uchida:1991}. We used the well-developed Kramers-Kronig analysis to obtain the phase of the reflection coefficient from the measured reflectance. For performing the Kramers-Kronig analysis, the measured spectrum should be extrapolated to zero and infinity \cite{wooten}. For the extrapolation to zero, we employed the Hagen-Rubens relation, i.e., $1-R(\omega) = \sqrt{2\omega \rho/\pi}$, using the measured DC resistivity data, $\rho$ (see Fig. \ref{fig0}). For the extrapolation to infinity, we used $R(\omega) \propto \omega^{-1}$ from 20,000 to $10^{6}$ cm$^{-1}$ and above $10^{6}$ cm$^{-1}$ we assumed the free electron behavior, i.e., $R(\omega) \propto \omega^{-4}$. We obtained the complex optical conductivity using the Fresnel formula and well-known relationships between the optical constants\cite{wooten,tanner:2019}.

\section{Results and discussion}

\begin{figure}[!htbp]
  \vspace*{-0.3 cm}%
  \centerline{\includegraphics[width=4.5 in]{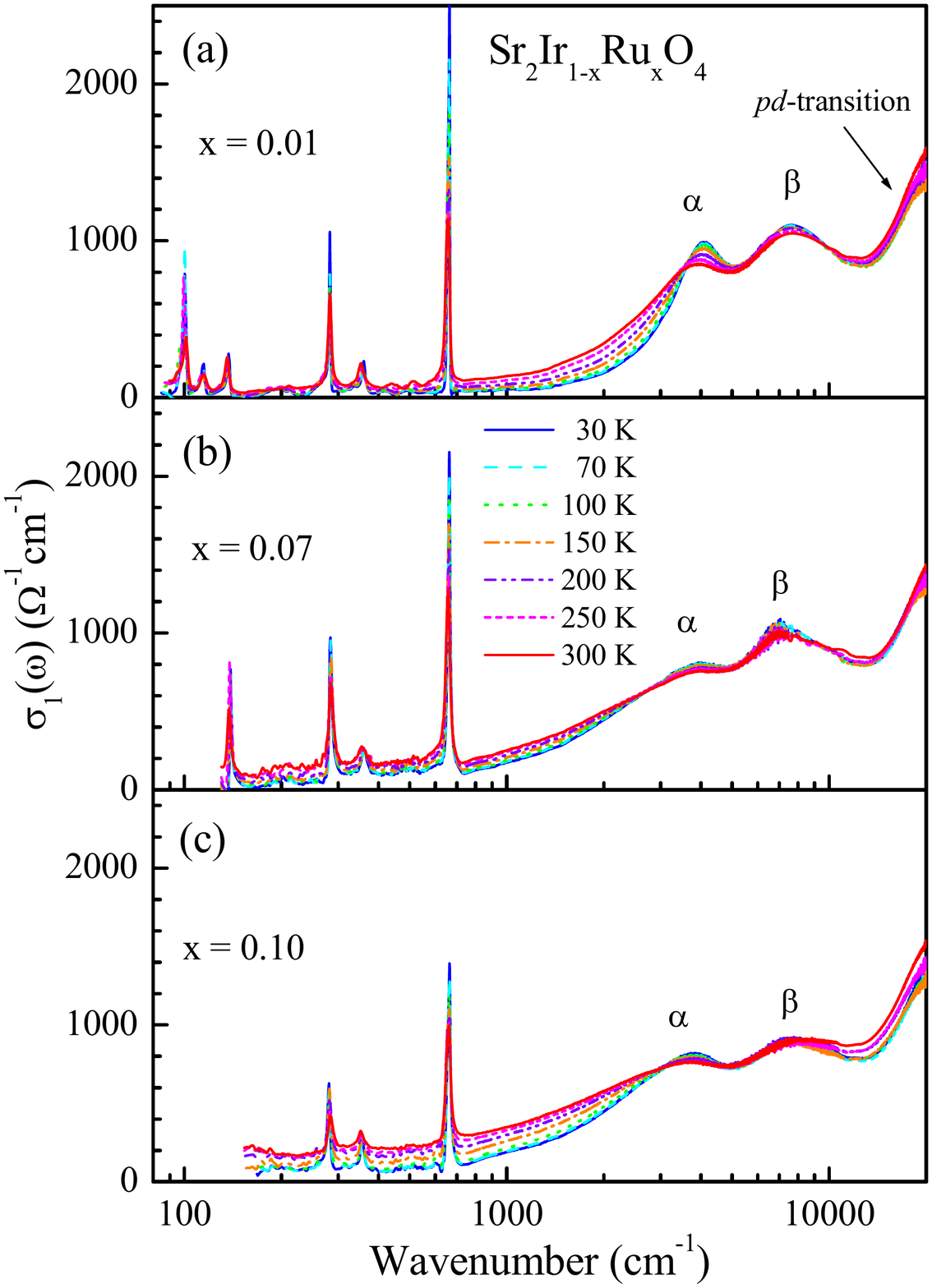}}%
  \vspace*{-0.7 cm}%
\caption{(Color online) Optical conductivity spectra of Sr$_2$Ir$_{1-x}$Ru$_x$O$_4$ in three different Ru-doping levels at various selected temperatures: (a) $x$ = 0.01, (b) $x$ = 0.07, and (c) $x$ = 0.10.}
 \label{fig2}
\end{figure}

Fig. \ref{fig2} shows the optical conductivity spectra of our three samples (Sr$_2$Ir$_{1-x}$Ru$_{x}$O$_4$: $x =$ 0.01, 0.07, and 0.10) at various selected temperatures from 30 to 300 K. The overall level of the optical conductivity in the high frequency region above 10,000 cm$^{-1}$ is slightly higher than that of undoped Sr-214 obtained using an ellipsometry\cite{sohn:2014}. It is worth to be noted that the optical conductivity in the high frequency region can be influenced by extrapolations to infinity for the Kramers-Kronig analysis. Below 10,000 cm$^{-1}$, we observed two prominent interband transitions, which were previously observed in undoped Sr-214\cite{kim:2008}. We also observed another interband transition above 15,000 cm$^{-1}$, which has been observed and assigned as a $pd$-transition\cite{sohn:2014}. We adopted the previous naming of the two prominent modes. We denoted them as $\alpha$ (near 4000 cm$^{-1}$) and $\beta$ (near 7500 cm$^{-1}$) modes. These modes appear owing to the strong spin-orbit-coupling ($E_{SO}$) in Sr-214, which splits the $t_{2g}$ orbital into $J_{eff}$ = 1/2 and 3/2 orbitals, and the Coulomb repulsion ($U$), which splits the half-filled $J_{eff}$ = 1/2 state into the lower Hubbard band (LHB) and upper Hubbard band (UHB)\cite{kim:2008,kuriyama:2010}. Therefore, the ground state of the undoped Sr-214 is insulating. The $\alpha$ mode has been identified as the optical transition from the occupied $J_{eff}$ = 1/2 LHB to the unoccupied $J_{eff}$ = 1/2 UHB and the $\beta$ mode as the transition from the occupied $J_{eff}$ = 3/2 band to the unoccupied $J_{eff}$ = 1/2 UHB. Undoped Sr-214 has been known as a $J_{eff}$ = 1/2 Mott insulator. In a previous study\cite{lee:2012}, measured optical spectra at room temperature of Ru-doped Sr-214 thin films exhibited strong Drude components, indicating that the  $J_{eff} =$ 1/2 LHB is partially filled by charge carriers introduced by Ru-doping. However, as shown in Fig. \ref{fig2}, the optical conductivity spectra of Ru-doped Sr-214 single crystals were significantly different from those in a previous optical study on Ru-doped Sr-212 thin films\cite{lee:2012}. In the previous study, a thin film with a continuous doping gradient was prepared on a substrate using a pulsed laser deposition; different spots in the film have different doping levels. The optical conductivity was estimated through a numerical subtraction of the substrate contribution using an approximate formula\cite{okazaki:2006}. The resulting conductivity spectra may contain unexpected uncertainty, which may result from the subtraction process. Additionally, the measured spectra might be affected by some extrinsic effects such as a lattice mismatch between the thin film and substrate. In fact, one earlier study demonstrated that the strain could be used to tune the electronic bandwidth, which caused changes in the overall electronic band diagram\cite{nicols:2013,bhandari:2019}. However, the causes of differences between the two data sets obtained from thin films and single crystals have not been clearly understood yet.

\begin{figure}[!htbp]
  \vspace*{-0.3 cm}%
  \centerline{\includegraphics[width=4.7 in]{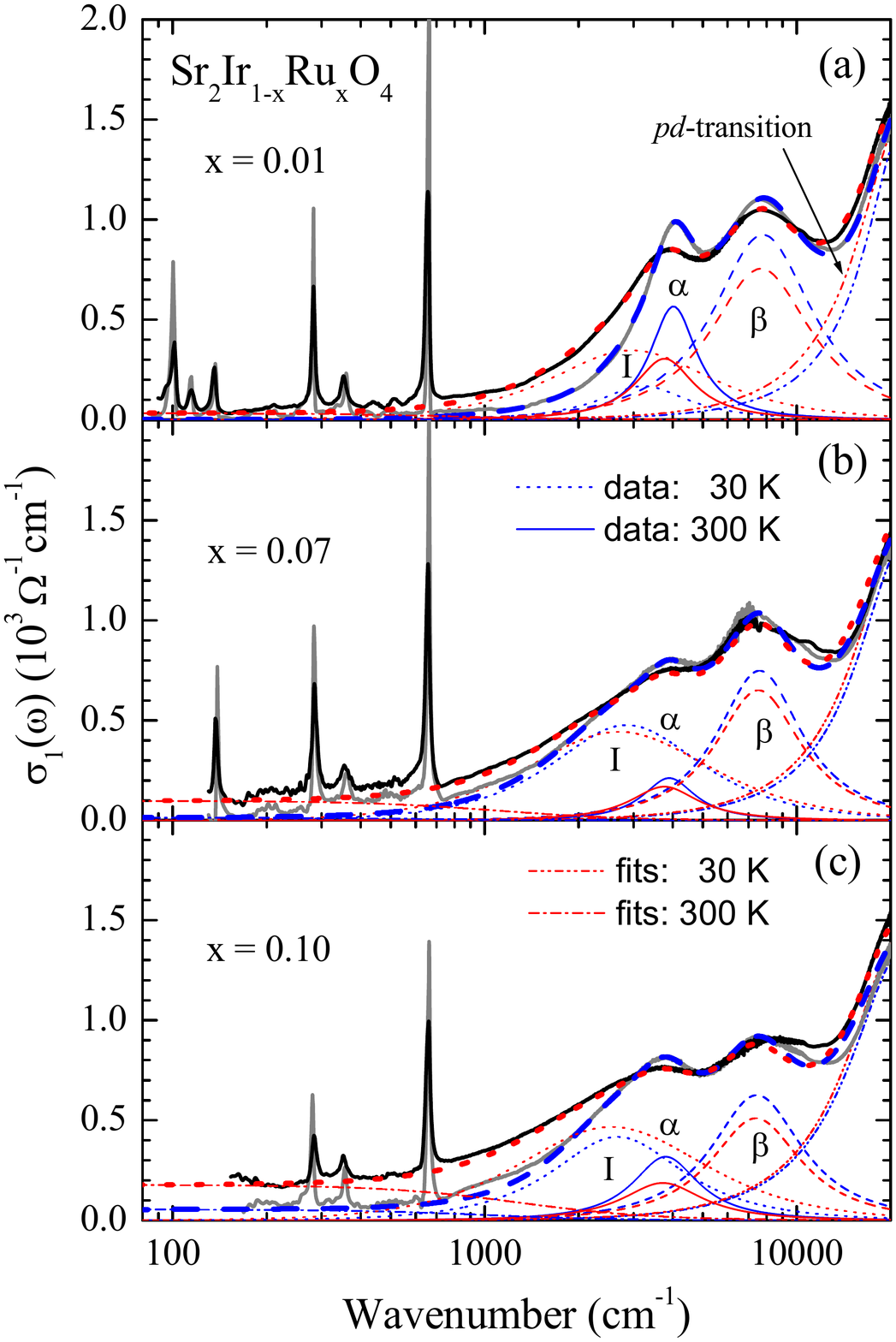}}%
  \vspace*{-0.5 cm}%
\caption{(Color online) Optical conductivity data and Drude-Lorentz model fits of Sr$_2$Ir$_{1-x}$Ru$_x$O$_4$ ($x$ = 0.01, 0.07, and 0.10) at two representative temperatures: 30 and 300 K. The dotted, solid, and dashed lines represent the inner-gap excitation (or the $I$ mode), $\alpha$ mode, and $\beta$ mode, respectively. The dot-dashed and dot-dot-dashed lines represent the Drude modes and $pd$-transitions, respectively.}
 \label{fig3}
\end{figure}

Fig. \ref{fig3} shows the optical conductivity spectra and Drude-Lorentz model fits of our three samples at two representative temperatures, 30 and 300 K. In the Drude-Lorentz model, the real part of the optical conductivity can be expressed as
\begin{equation}\label{eq1}
\sigma_1(\omega)=\frac{\Omega_{D,p}^2}{4\pi}\frac{1/\tau_{imp}}{\omega^2+(1/\tau_{imp})^2} + \sum_k\frac{\Omega_{k,p}^2}{4\pi}\frac{\gamma_k\omega^2}{(\omega_k^2-\omega^2)^2+\gamma_k^2\omega^2},
\end{equation}
where $\Omega_{D,p}$ is the Drude plasma frequency and $1/\tau_{imp}$ is the elastic impurity scattering rate. $\omega_k$, $\Omega_{k,p}^2$, and $\gamma_k$ are the center frequency, strength, and width (or damping parameter) of the $k$th Lorentz oscillator, respectively. Ru-doping introduces another new mode below the $\alpha$ mode, which was observed and interpreted as an inner-gap excitation in previous optical studies on Ru-doped Sr-214 thin films\cite{lee:2012} and La-doped Sr-214 single crystal samples\cite{seo:2017,wang:2018}. For the fitting, we used one Drude mode, three ($I$, $\alpha$, and $\beta$) Lorentz modes, and one additional interband transition located at the high frequency region, known as a $pd$-transition\cite{sohn:2014}. Here, we denote the inner-gap excitation as the $I$ mode. Similar inner-gap excitations have been observed in 3$d$ transition metal oxides when they are doped\cite{imada:1998}. This new mode may be related to the famous mid-infrared absorption in cuprates\cite{uchida:1991,quijada:1999}. The absorption can be explained as the incoherent component of the spectral weight of charge carriers induced by a strong correlation\cite{hwang:2008a}. However, the origin of the inner-gap excitation in Ru-doped Sr-214 systems is not clearly known yet. We separately show the Drude mode, the three prominent Lorentz modes, and the $pd$-transition in Fig. \ref{fig3}. We ignored the sharp phononic absorption peaks located in the low-frequency region below 800 cm$^{-1}$ for these fittings.

\begin{figure}[!htbp]
  \vspace*{-0.4 cm}%
  \centerline{\includegraphics[width=4.7 in]{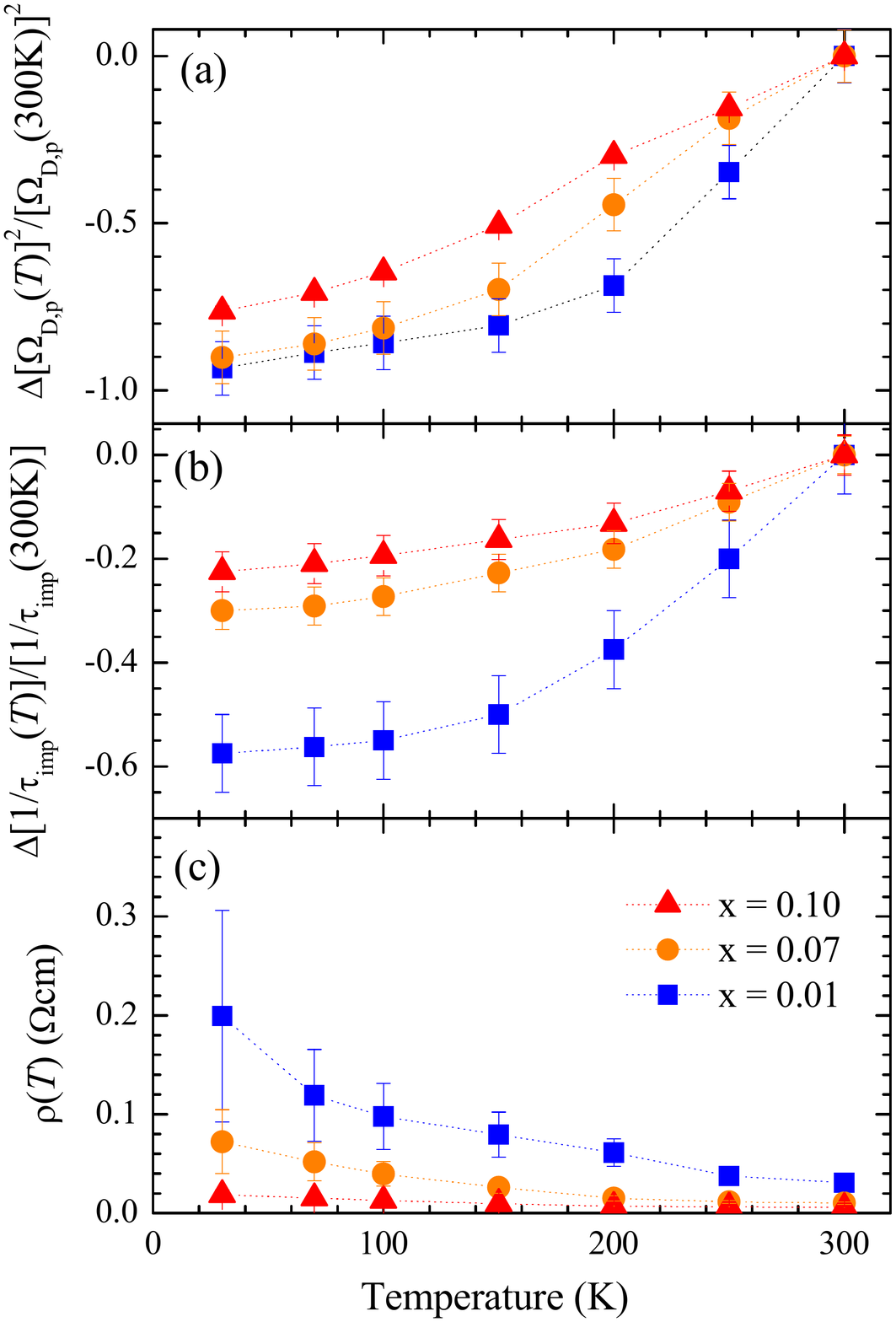}}%
  \vspace*{-0.5 cm}%
\caption{(Color online) (a,b) Normalized relative temperature-dependencies, $\Delta [\Omega_{D,p}(T)]^2/[\Omega_{D,p}(300K)]^2$ and $\Delta [1/\tau_{imp}(T)]/[1/\tau_{imp}(300K)]$ of the two Drude fitting parameters ($\Omega_{D,p}$ and $1/\tau_{imp}$)). (c) Temperature-dependent DC resistivity obtained from the Drude fitting parameters.}
 \label{fig4}
\end{figure}

In Fig. \ref{fig4}(a) and \ref{fig4}(b), we show normalized relative temperature-dependencies, $\Delta [\Omega_{D,p}(T)]^2/[\Omega_{D,p}(300 K)]^2$ and $\Delta [1/\tau_{imp}(T)]/[\tau_{imp}(300 K)]$, of two fitting parameters (plasma frequency and impurity scattering rate) of the Drude mode for the three Ru-doped samples ($x =$ 0.01, 0.07, and 0.10). The normalized relative temperature-dependencies are defined by $\Delta [\Omega_{D,p}(T)]^2/[\Omega_{D,p}(300 K)]^2 \equiv \{[\Omega_{D,p}(T)]^2-[\Omega_{D,p}(300 K)]^2\}/[\Omega_{D,p}(300 K)]^2$ and $\Delta 1/\tau_{imp}(T)/[1/\tau_{imp}(300 K)] \equiv [1/\tau_{imp}(T)-1/\tau_{imp}(300 K)]/[1/\tau_{imp}(300 K)]$. Pairs of $\Omega_{D,p}(300 K)$ and $1/\tau_{imp}(300 K)$ are 1250 and 800, 2550 and 1100, and 3700 and 1290 cm$^{-1}$, for the three samples with $x$ = 0.01, 0.07, and 0.10, respectively. The Drude plasma frequencies of all three samples exhibit significant temperature dependencies. The plasma frequency squared, $\Omega_{D,P}^2$, which is proportional to the charge-carrier density, monotonically increases with raising the temperature, probably due to thermal excitations. In general, the plasma frequency of a good metal is independent of temperature. The impurity scattering rates of all three samples decrease with lowering the temperature, which is a generic temperature-dependent trend of the impurity scattering rate of a good metal. Both plasma frequency and impurity scattering rate increase with the Ru-doping. Using these two Drude fitting parameters, we estimated dc resistivity ($\rho$), i.e., $\rho = 4\pi(1/\tau_{imp})/\Omega_{D,p}^2$. The estimated dc resistivity data of the three samples are shown as functions of temperature in Fig. \ref{fig4}(c). The obtained dc resistivity data of all three samples exhibit insulating temperature-dependent behavior, which is consistent with the directly measured DC resistivity data shown in Fig. \ref{fig0}. We can see huge discrepancy in low temperatures between the two sets of DC resistivity data. At low temperatures, the Drude component is very (or almost too) weak to get reasonable fitting parameters. Therefore, the very weak Drude component may cause the huge discrepancy between the two sets of the DC resistivity data at low temperatures.

\begin{figure}[!htbp]
  \vspace*{-0.3 cm}%
  \centerline{\includegraphics[width=4.7 in]{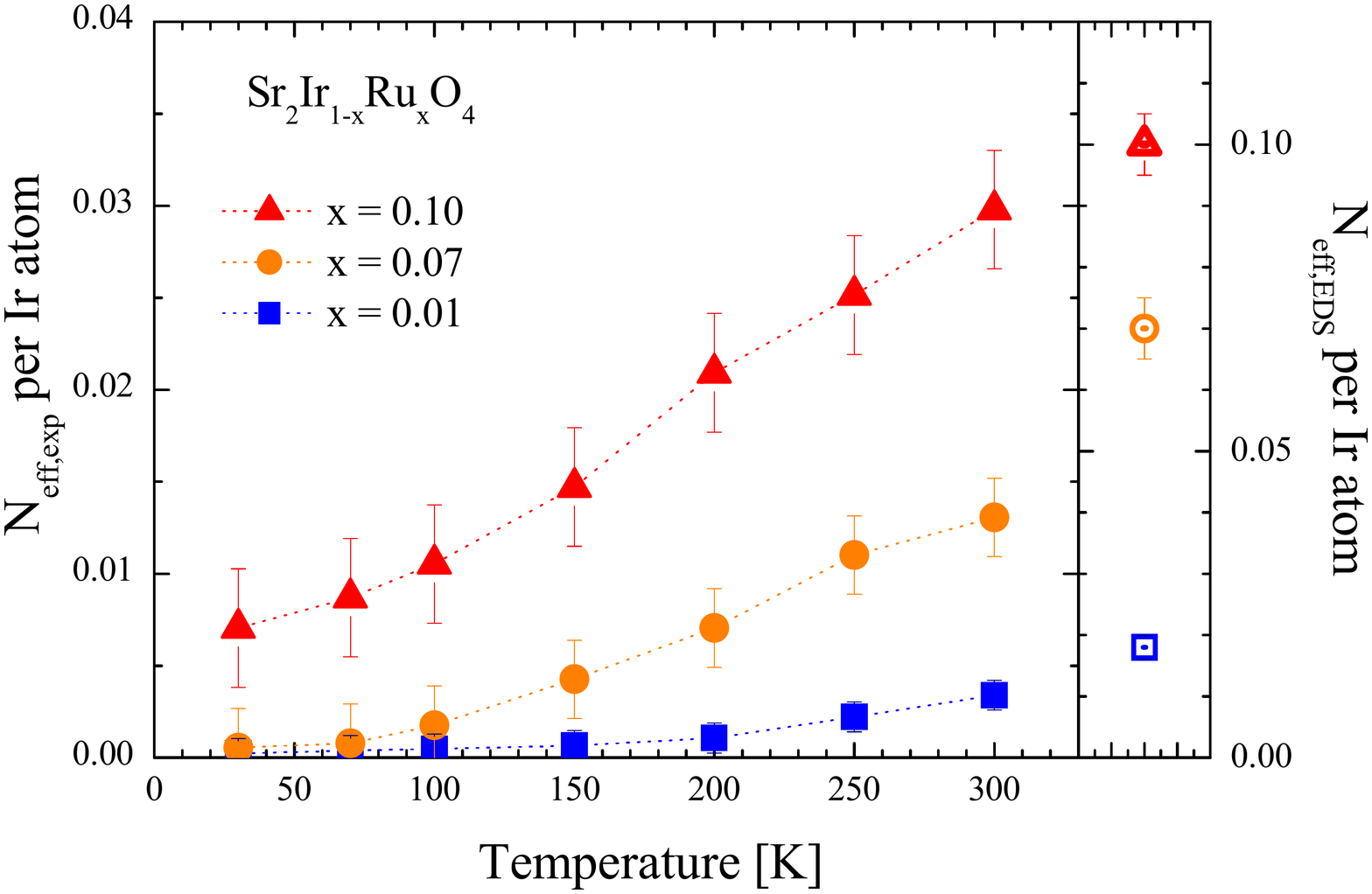}}%
  \vspace*{-0.5 cm}%
\caption{(Color online) Experimentally estimated number of charge-carriers per Ir atom of the three samples (Sr$_2$Ir$_{1-x}$Ru$_x$O$_4$; $x$ = 0.01, 0.07, and 0.10) as functions of temperature (in the main frame) and the charge-carriers per Ir atom for the three samples (in the narrow right frame) estimated using the EDS technique.}
 \label{fig5}
\end{figure}

Fig. \ref{fig5} shows the effective charge-carrier number per Ir atom, $N_{eff, exp}$, of the three Ru-doped Sr-214 samples obtained from the Drude plasma frequencies. The the effective charge-carrier number per Ir atom ($N_{eff, exp}$) can be estimated using the relationship between the plasma frequency and carrier density, i.e., $N_{eff, exp} = \Omega_{D,p}^2 m_e V_{Ir}/(4\pi e^2)$, where $m_e$ is the electron mass, $e$ the elementary charge, and $V_{Ir}$ the volume per one formula unit. The experimentally estimated $N_{eff,exp}$ exhibited strong Ru-doping and temperature dependencies. As the temperature increased, $N_{eff, exp}$ significantly increased due to the thermal excitation effect, which are closely related to the small activation energies estimated from the DC resistivity data (see the inset of Fig. \ref{fig0}). The estimated $N_{eff, exp}$ were significantly smaller than the expected ones ($N_{eff, EDS}$), which were estimated using the EDS technique. The $N_{eff,EDS}$ estimated by the EDS are shown in the narrow frame on the right in Fig. \ref{fig5}, where the vertical axis scale is around 3 times larger than that of the main frame. This kind of discrepancy between the estimated values from the optical spectroscopy and EDS technique can be observed between two infrared studies of La-doped Sr-214 system (Sr$_{2-y}$La$_y$IrO$_4$)\cite{seo:2017,wang:2018}. The related problem of doping may be seen between two samples (Sr$_{2-y}$La$_y$IrO$_4$ with $y$ = 0.1 and 0.18) in an optical study\cite{wang:2018}. It is not clear why the estimated $N_{eff,exp}$ is significantly smaller than the EDS one. A possible interpretation of this low charge-carrier density may be associated with the effective mass. For the estimation, we considered the mass of charge carriers as the bare electron mass; however, the mass can be enhanced owing to the correlations. Typically, the spectral weight is divided into two components (coherent and incoherent) owing to the correlation\cite{hwang:2008a,basov:2011a}. Here, the Drude spectral weight ($\propto N_{eff,exp}$) can be approximately that of the coherent component, and the total spectra weight can be the EDS spectral weight ($\propto N_{eff,EDS}$). With these values at 30 K, we obtained the coupling constants ($\lambda$), which can be defined as $\lambda +1 \equiv N_{eff,EDS}/N_{eff,exp}$. The coupling constants of the three samples are 459, 1287, and 14.2 for the samples with $x =$ 0.01, 0.07, and 0.10, respectively), which seemed to be way too large. Hence, other unknown doping-related phenomena may exist in Ru-doped Sr-214, as in the Rh-doped Sr-214\cite{clancy:2014,cao:2016}.

\begin{figure}[!htbp]
  \vspace*{-0.4 cm}%
  \centerline{\includegraphics[width=6.5 in]{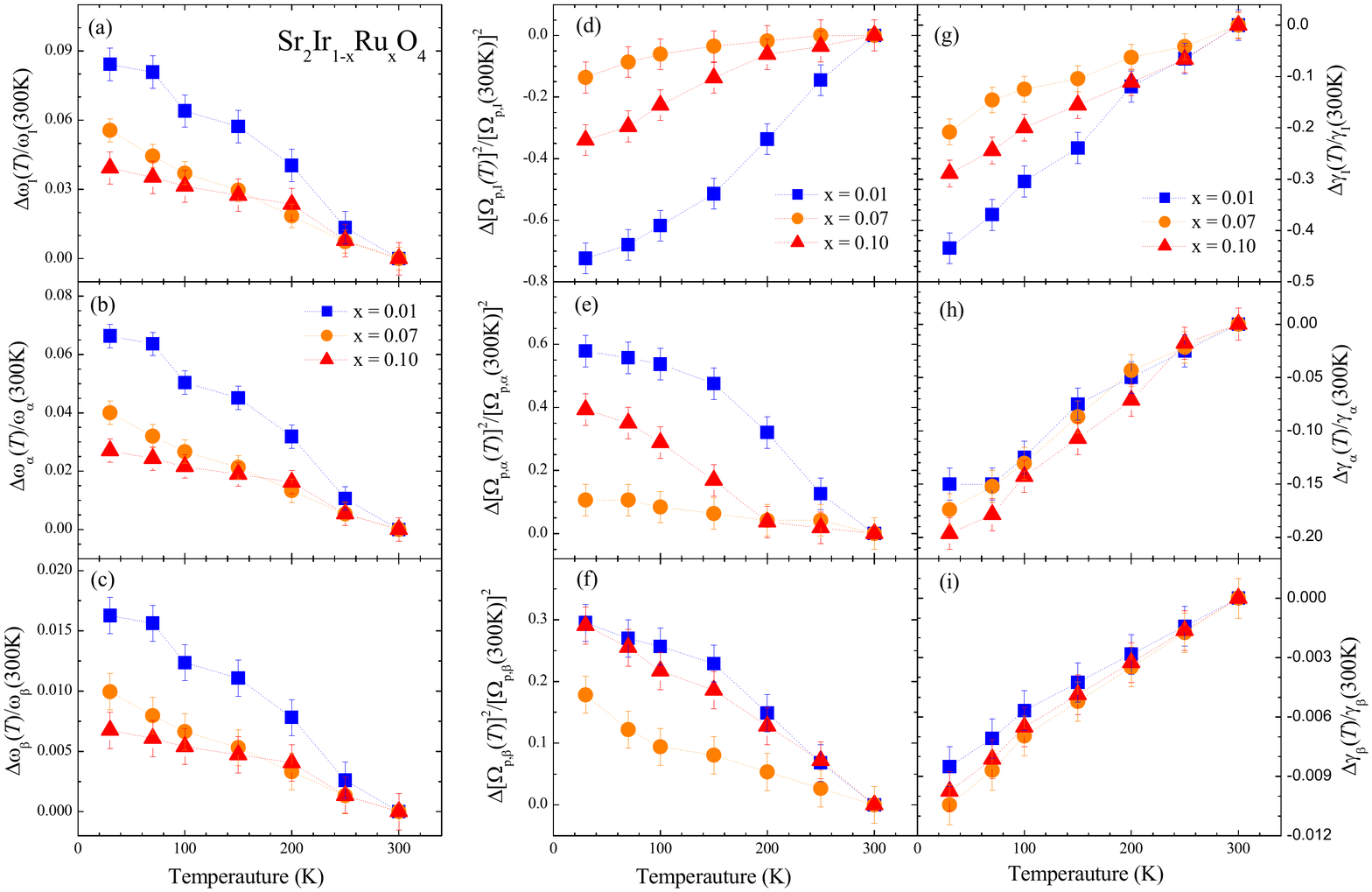}}%
  \vspace*{-0.5 cm}%
\caption{(Color online) Normalized relative temperature-dependent center frequencies (a-c), strength ($\Omega_p^2$) (d-f), and damping parameters (g-i) of the inner-gap excitation ($I$ mode), $\alpha$, and $\beta$ modes for the three Ru-doped Sr-214 samples.}
 \label{fig6}
\end{figure}

Fig. \ref{fig6} shows the doping- and temperature-dependent fitting parameters of three Lorentz ($I$, $\alpha$, and $\beta$) modes for the three samples. We took a relative value to the value at 300 K and normalized with the value at 300 K, i.e., $\Delta A(T)/A(300 K) \equiv [A(T)-A(300 K)]/A(300 K)$, where $A$ is a representative quantity. The temperature-dependent properties of the $\alpha$ and $\beta$ modes of undoped Sr-214 have been previously reported in the literature\cite{moon:2009,sohn:2014}. Our sample at the lowest doping concentration exhibited a similar temperature-dependence as that reported previously. As the temperature increases, the $\alpha$ mode shifts to a lower frequency. The temperature-dependent trend remains the same as that of undoped Sr-214 (Fig. \ref{fig6}(b)). As the Ru-doping increases, the $I$ mode shifts to a lower frequency as well (see Table I). The $\alpha$ and $\beta$ modes exhibit similar doping- and temperature-dependent trends as the $I$ mode. Similar doping-dependent shifts of the $\alpha$ and $\beta$ modes were observed in a recent optical study on La-doped Sr-214\cite{seo:2017}. We also observe that the temperature-dependent shift of the $\alpha$ mode ($\Delta \omega_{\alpha}(T) \equiv \omega_{\alpha}(T)-\omega_{\alpha}(300K)$) is twice that of $\beta$ mode ($\Delta \omega_{\beta}(T) \equiv \omega_{\beta}(T)-\omega_{\beta}(300K)$), as reported in a previous paper\cite{sohn:2014}, i.e.,$\Delta \omega_{\alpha}(T) = 2\Delta \omega_{\beta}(T)$. A previous study proposed that the observed shift ratio between the two interband ($\alpha$ and $\beta$) transition energies is associated with the electron-phonon coupling\cite{sohn:2014}. This energy shift ratio between the $\alpha$ and $\beta$ modes is not dependent on Ru-doping. We note that $2\omega_{\alpha}(300K) \cong \omega_{\beta}(300K)$ (see Table I). At each temperature, all three modes shifts to a lower frequency as the doping increases (see Table I). The normalized relative temperature-dependent strength ($\Omega_p^2$) (d-f) and damping parameters (g-i) of the three Lorentz modes are shown in the right two columns in Fig. \ref{fig6}. We also provide all fitting parameters of the three Lorentz modes at 300 K in Table I. We observe systematic temperature-dependent trends of the strength and damping parameter. However, as we can see in Table I, the strength of the $I$ mode increases with the Ru-doping, while those of both $\alpha$ and $\beta$ modes decrease, which are similar doping-dependent trends in the strengths of the three Lorentz modes observed in La-doped Sr-214 systems\cite{seo:2017}. With lowering the temperature, the intensity of $I$ mode decreases, while those of both $\alpha$ and $\beta$ modes increase, indicating that the spectral weight of the $I$ mode shifts to the $\alpha$ and $\beta$ modes. All three damping parameters monotonically decrease with lowering the temperature, which is a general temperature-dependent trend of the damping parameter.

\begin{table}[h!]
\begin{tabular}{|c|c|c|c|c|c|c|c|c|c|}
  \hline\hline
   samples & $\omega_I$ & $\omega_{\alpha}$ & $\omega_{\beta}$ & $\Omega_{p,I}$ & $\Omega_{p,\alpha}$ & $\Omega_{p, \beta}$ & $\gamma_I$ & $\gamma_{\alpha}$ & $\gamma_{\beta}$ \\
  \hline\hline
  $x$ = 0.01 & 2970 & 3770 & 7675 & 9700 & 6050 & 17350 & 4600 & 2000 & 7060 \\
  \hline
   0.07 & 2700 & 3750 & 7530 & 11300 & 4850 & 15200 & 4800 & 2300 & 5750 \\
  \hline
   0.10 & 2550 & 3700 & 7400 & 11200 & 5550 & 12662 & 4500 & 2800 &6150 \\
  \hline\hline
\end{tabular}
\caption{All fitting parameters of the three ($I$, $\alpha$, and $\beta$) Lorentz modes at 300 K. Note that units for the all values are cm$^{-1}$.}
\label{table:1}
\end{table}

\begin{figure}[!htbp]
  \vspace*{-0.4 cm}%
  \centerline{\includegraphics[width=6.5 in]{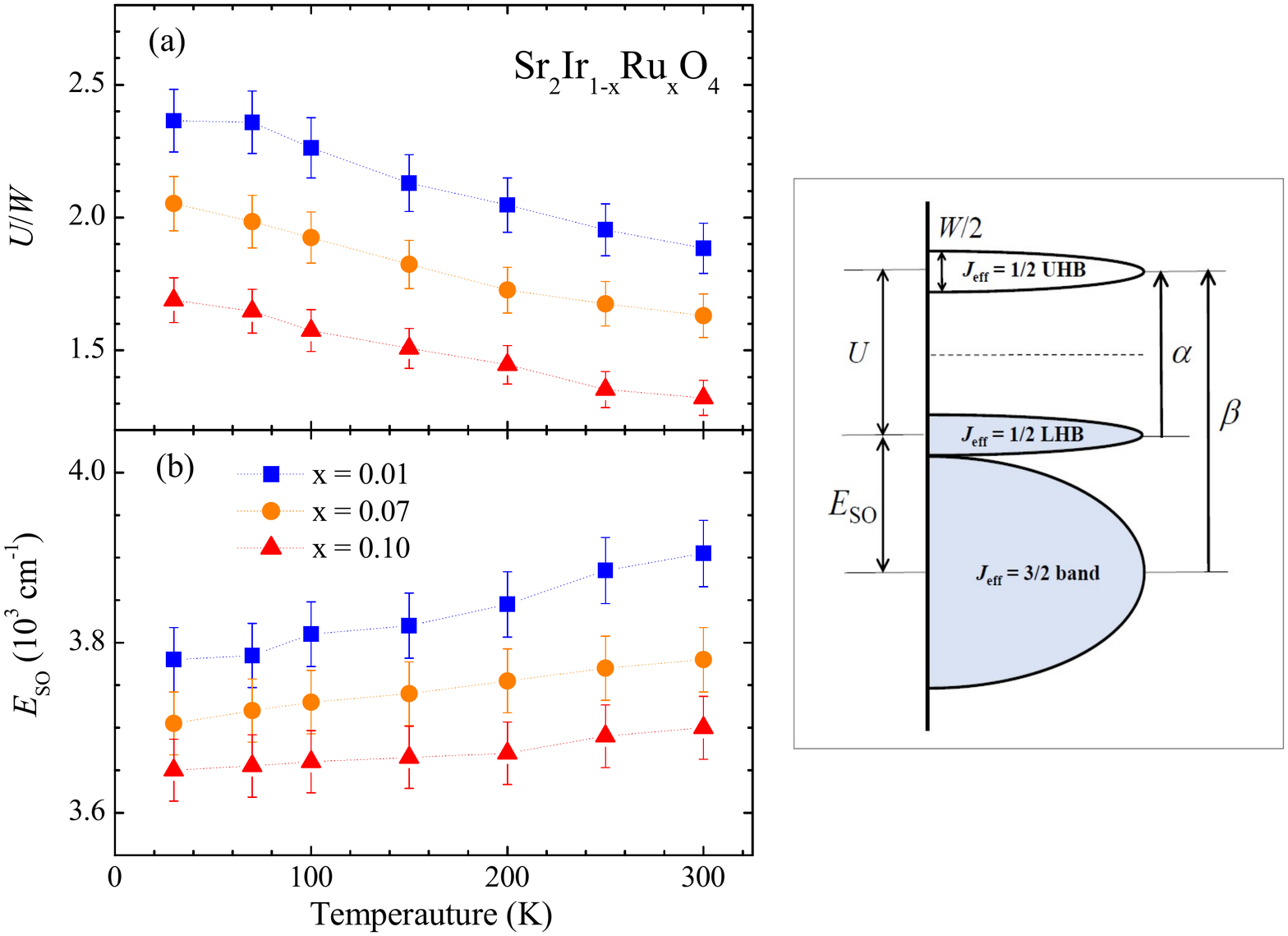}}%
  \vspace*{-0.5 cm}%
\caption{(Color online) Temperature-dependent (a) ratio of the Coulomb repulsion to the bandwidth ($U/W$) and (b) spin-orbit coupling energy ($E_{SO}$) for the three Ru-doped Sr-214 samples as functions of temperature. On the right, a schematic band diagram is shown.}
 \label{fig7}
\end{figure}

Fig. \ref{fig7} shows the temperature-dependent ratio of the Coulomb repulsion to the bandwidth ($U/W$) and the spin-orbit coupling energy ($E_{OS}$) for the three samples. As shown in the schematic band diagram in the right figure, $U$ is the same as the center frequency of the $\alpha$ mode and $W$ is the same as the width of the $\alpha$ mode. $E_{SO}$ is obtained from the center frequencies of $\alpha$ and $\beta$ modes, i.e., $E_{SO} = \omega_{\beta} - \omega_{\alpha}$. The schematic band diagram is based on our optical results. As the temperature increases, $U/W$ decrease because the thermal broadening of the $\alpha$ mode makes $W$ larger and the shifting to lower energy of the $\alpha$ mode makes $U$ smaller (see Fig. \ref{fig6}(b) and (h)). As the doping increases, $U/W$ also significantly decreases because the center frequency of the $\alpha$ mode decreases while its width increases with increasing the doping as shown in Table I. As the temperature increases, $E_{SO}$ increases because the temperature-dependent changes between the $\alpha$ and $\beta$ modes is different; the center frequency of the $\beta$ mode decreases less than that of the $\alpha$ mode as we discussed in the previous paragraph, i.e., $\Delta \omega_{\alpha}(T) = 2\Delta \omega_{\beta}(T)$. As doping increases, $E_{SO}$ decreases because the center frequency of the $\beta$ mode decreases more than that of the $\alpha$ mode with increasing the doping as shown in Table I, which is consistent with a previously reported one\cite{lee:2012}.

\section{Conclusions}

We measured reflectance spectra of three differently Ru-doped Sr-214 single crystal samples. We obtained the optical conductivity spectra of the three samples from the measured reflectance spectra using the Kramers-Kronig analysis. We fitted the optical conductivity spectrum with the well-known Drude-Lorentz model. The charge-carrier number per Ir atom estimated from the Drdue plasma frequency was significantly smaller than the expected value estimated from the nominal doping concentration. The origin of low charge-carrier number has not yet been understood. All three samples exhibited insulating temperature-dependent behavior, indicating that the electronic ground state is insulating. Three major interband optical transitions ($I$, $\alpha$, and $\beta$) were observed. From the Drude-Lorentz fitting parameters, we obtained temperature- and Ru-doping-dependent $U/W$ and $E_{SO}$. With increasing temperature, the $U/W$ decreased while $E_{SO}$ increased. With increasing doping, both $U/W$ and $E_{SO}$ decreased. We hope that our experimental results will be useful for understanding the physical properties of Ru-doped Sr-214 systems.

%
%
\acknowledgments J.H. acknowledges financial support from the National Research Foundation of Korea (NRF-2017R1A2B4007387 and 2021R1A2C101109811). The work at Yonsei University was supported by the National Research Foundation of Korea (grant numbers NRF-2017R1A5A1014862 (SRC program: vdWMRC center), NRF-2019R1A2C2002601, and NRF-2021R1A2C1006375).

%
%

\bibliographystyle{apsrev4-1}
\bibliography{bib}

\end{document}